Configurations of the third nearest-neighbor molecules forming a vacancy wall

and an addition of a $CO_2$ molecule in the vacancy

of solid $CO_2$ at $T$ = 0, 100, and 200 K studied by Monte Carlo simulation technique

(*Second version revised on October 20, 2020*)

Koji Kobashi

Former Research Assistant, Physics Department, Colorado State University, Fort Collins, CO, USA,

and

Former Senior Researcher, Kobe Steel, Ltd., Japan

Abstract

Configurations of the molecules on the wall of a vacancy, formed by removing a central and its first and second nearest-neighbor (NN) molecules in solid $CO_2$ with the *Pa3* structure, were calculated by the Monte Carlo simulation technique at $T$ = 0, 100, and 200 K and a nominal pressure of $P$ = 1 bar. It was found that the deviations of both the center-of-mass and the orientational coordinates of the molecules from the unperturbed coordinates had a three-fold symmetry about a body diagonal axis of the crystal. It was also found that a single $CO_2$ molecule, initially placed in the center of the vacancy, was stabilized at a position close to the vacancy wall. This paper is a continuation of arXiv:1711.04976 [cond-mat.mtrl-sci] (2017) and arXiv:1809.04291 [cond-mat.mtrl-sci] (2018).

1. Introduction

The present work is a continuation of the precedent two articles[1,2] that investigated molecular configurations surrounding a single vacancy in solid $CO_2$ with the *Pa3* structure[3,4] at temperatures of $T$ = 0, 100, and 200 K using the Monte Caro (MC) simulation technique[5] and a Kihara core potential model.[6] In the articles, the basic cell was cubic, and contained 2047 (=2048-1) molecules, to which the periodic boundary condition was applied. It was found that at $T$ = 0 and 100 K, the orientational changes were greater for the three nearest-neighbor (NN) molecules and the equivalent three other molecules located close to the oxygen atoms associated with a $CO_2$ molecule that had been removed to make a vacancy at the origin. By contrast, such changes disappeared at $T$ = 200 K as the molecular orientations were randomized due to high temperature. In the first part of the present work, a large vacancy was made by removing a molecule in the center and its first and second NN molecules: namely, 19 molecules were removed to make a cubic vacancy. Hence, the third NNs consisting of 24 molecules form a wall of the vacancy, and these molecules will be hereafter referred to as "wall molecules". Thus, the total number of molecules in the cubic basic cell with the vacancy was 2029 (=2048-19). For reference, simulations were also done for $CO_2$ crystals without vacancy. In the second part of the paper, a stable configuration of a $CO_2$ molecule that was initially placed at the origin of the vacancy was studied. Since solid $CO_2$ sublimes at $T$ = 194.7 K at the ambient pressure, one may expect that the $CO_2$ molecule moves freely in the vacancy at $T$ = 200 K. The Monte Carlo simulations showed, however, that the $CO_2$ molecule stayed close to the wall molecules. The major interest of the present work is to know the influence of temperature on the configurations of the 24 wall molecules and the added $CO_2$ molecule.

2. Computational procedure

The computational procedure was similar to those used in Refs. 1 and 2. The lattice constants $a$ used at different temperatures were assumed in reference to experiments[6,7] as follows: $a$ = 5.53725 + 4.679 $\times 10^{-6}$ $T^2$ in units of Å. The edge of the cubic basic cell consisted of eight primitive unit cells so that the length was 44.3 Å at $T$ = 0 K. Note, therefore, that the vacancies form a cubic lattice with a lattice constant of 44.3 Å at $T$ = 0 K that is far enough for vacancies to interact with each other because the maximum interaction range between molecules was set to be 15 Å in the present work. The same applies to the cases of $T$ = 100 and 200 K. The periodic boundary condition was applied to the cubic basic cell. The pressure was not calculated in the present work but likely to be ambient as the temperature dependence of $a$ was determined from the experiments at ambient pressure as described above. In the first part of the present paper, there were 2029 molecules in the cubic basic cell containing a cubic vacancy that was formed by removing 19 molecules. The initial crystal structure was $Pa3$, and both the center-of-mass (CM) and orientational coordinates of all molecules in the basic cell were randomized within a range of -0.2 Å < $\Delta x, \Delta y, \Delta z$ < 0.2 Å (1 Å = 0.1 nm) for CM positions and within a range of -30° < $\Delta\theta, \Delta\phi$ < 30° for molecular orientations. These are collectively expressed as {$\Delta r, \Delta\psi$} = {0.2 Å, 30°}, hereafter. The randomized structure returned to a structure close to $Pa3$ in the next computational job as indicated by the CM positions, the molecular orientations as well as the total crystal energy. The values of {$\Delta r, \Delta\psi$} were then decreased stepwise to {0.1 Å, 10°} in the default cases. The last 40 data were chosen for analysis after the total energy had been stable. A single computational job contained 100,000 rounds of MC calculations over all molecules in a random order. The acceptance ratio was only $5\times10^{-3}$ %. The major reason for the extremely low acceptance ratio is attributable to the large values for {$\Delta r, \Delta\psi$} as new molecular configurations are mostly rejected, but the actual reason is still under investigation. The computations were done on regular desktop computers using gfortran, a Linux Fortran. Each job took about 130 minutes. The total energy of the $CO_2$ crystal reached an equilibrium approximately after 20 jobs from the start. In the second part of the present paper, a $CO_2$ molecule oriented in the (111)-direction was initially placed at the origin of the vacancy, and the MC simulations were performed at the three different temperatures to determine the stable CM position and the orientation of the molecule. Note also that solid $CO_2$ (dry ice) sublimes at 194.7 K at ambient pressure, while solid $CO_2$ at 200 K was simulated in the present work. This does not seem to affect the major results in the present work as the potential model used does not have such a high accuracy as to quantitatively reproduce the sublimation temperature though it needs to be proved. It should be noted that unlike regular MC simulations, the energy calculations in the present paper, and in Refs. 1 and 2 as well, were carried out not by randomly choosing a molecule in the cubic basic cell in every calculation but by choosing all molecules in the basic cell in a random sequence for each calculation. This protocol seemed to be most efficient for the convergence of the crystal energy.

## 3. Results and Discussion

### 3.1 Wall molecules in the vacancy

Figure 1 shows the configurations of the wall molecules at $T = 0$ K observed from the [111] direction of the *Pa3* crystal. The molecules are numbered from 1 to 24. The CM of each molecule is closely located on an *fcc* lattice site, and closely oriented along one of the body diagonals of the cubic lattice. A brown sphere and two red spheres in a molecule indicate carbon and oxygen atoms, respectively, with van der Waals radii.

Table 1 shows the calculated data at $T = 200$ K for explaining one of the major results of the present work. In the table, the notations are as follows:

Mol. No. : molecular label shown in Fig. 1,

Position : approximate direction of the molecule seen from the origin of the vacancy,

Orientation : orientation of the molecule in the *Pa3* structure,

$\delta x$, $\delta y$, and $\delta z$ : CM deviations in the *x*-, *y*-, and *z*-directions in units of Å from the initial positions of the *Pa3* structure, respectively,

$\delta p$, $\delta q$, and $\delta r$ : deviations of the unit vector along the molecular axis in the *x*-, *y*-, and *z*-directions in units of Å from the initial orientation of the *Pa3* structure, respectively, and

$\delta E$ : molecular energy in units of K (1 K = $1.38054 \times 10^{-23}$ J) measured from the molecular energy in the unperturbed crystal structure with the vacancy.

Note in Fig. 1 that there are three molecules forming a triangle at each corner of the cubic vacancy, and in Table 1, the three molecules are shown in different colors. In Table 1, it is seen that the CM and orientational deviations have a three-fold symmetry about the [111] direction of the crystal: the combinations are 2-5-13, 4-11-14, 6-10-17, 7-16-20, 8-15-19, and 12-18-21 in terms of Molecular Number (Label). In each of the combinations, the molecular energies $\delta E$ are also very close. The three-fold symmetry was also present for $T = 0$ and 100 K cases.

Table 2 shows the averaged deviations of the CM positions $<r>$ and the orientations $<\Omega>$ as well as the averaged standard deviations (STDVs) of the CM positions $<<r>>$ and the orientations $<<\Omega>>$ over the wall molecules at different temperatures in reference to the CM positions and the orientations in the unperturbed *Pa3* structure at the corresponding temperature. Note that $<r>$ and $<\Omega>$ are absolute values. The results in Table 2 are to be compared with those of a $CO_2$ crystal without vacancy shown in Table 3. Since the wall molecules have an empty space on one side, all values of $<r>$, $<\Omega>$, $<<r>>$, and $<<\Omega>>$ in Table 2 are greater than those in Table 3. The results in Table 3 indicates that the cubic vacancy surrounded by the wall molecules is quite robust even at high temperature in a sense that the wall molecules change their CM positions only slightly even at $T = 200$ K.

3.2 A single molecule in the vacancy

Figure 2 shows the configurations of the wall molecules and the added single molecule at $T = 200$ K observed from the [111] direction of the *Pa3* crystal. The single molecule was initially placed at the origin of the cubic vacancy with the (111) orientation, but in equilibrium, it moved to a position close to the molecular clusters 2-4-10 (in the <11-1> direction from the origin) and 7-8-12 (in the <1-1-1> direction from the origin). The CM position of the single molecule was close to a position of the first NN molecule that was located in the <10-1> direction from the origin, corresponding to molecule 32 of Ref. 1. The molecular orientation was $\{p, q, r\} = \{0.27, 0.67, -0.67\}$ that was significantly different from $\{0.58, 0.58, -0.58\}$ for molecule 32 of Ref. 1. By contrast, the CM position and the orientation were similar to those of a corresponding NN molecule for the case of $T = 0$ and 100 K, and the configuration of the single $CO_2$ molecule was less randomized than at $T = 200$ K.

Table 4 shows the deviations of CMs, orientations, and molecular energies of the wall molecules at $T = 200$ K from the corresponding values in the *Pa3* structure. The notations are the same as in Table 1. Unlike in Table 1, the three-fold symmetry was not observed among the wall molecules, most likely because the added $CO_2$ molecule perturbed the configurations of the wall molecules. In fact, the energy deviations $\delta E$ are more negative than the counterparts shown in Table 1 because of the interactions with the added $CO_2$ molecule. Since the added $CO_2$ molecule is located closer to the molecular clusters 2-4-10 and 7-8-12, the molecular energies of molecules 2, 7, 10, and 12 are significantly lower than other wall molecules.

Table 5 shows averaged deviations of the CM positions $<r>$ and the orientations $<\Omega>$ as well as averaged STDVs of the CM positions $<<r>>$ and the orientations $<<\Omega>>$ over the wall molecules forming a vacancy including a single $CO_2$ molecule. All values tend to increase at higher temperature, but the changes were much smaller than the values expected before the present work. Table 6 shows averaged STDVs of the CM positions $<<r>>$ and the orientations $<<\Omega>>$ of the added single $CO_2$ molecule. Similarly, all values tend to increase at higher temperature, and the changes were more strongly dependent on temperature than for the wall molecules presumably because the added $CO_2$ molecule is less constrained than the wall molecules.

4. Conclusion

In the first part of the present article, it was found that the average molecular configurations of the wall molecules comprising the vacancy had a three-fold symmetry about a [111] body-diagonal axis of the crystal. The cubic vacancy structure was basically maintained even at $T = 200$ K, and both the deviations and the STDVs of the CMs positions and orientations for the wall molecules were small. This indicates that the cubic vacancy is quite robust to temperature. In the second part of the present article, it was found that a $CO_2$ molecule added to the vacancy stayed near the wall molecules. For a theoretically consistent study, it is necessary to undertake the pressure-constant MC simulations, but that is an issue to be investigated in the future.

*Acknowledgement*

Figures 1 and 2 are depicted using an open software, VESTA, published in: K. Momma and F. Izumi, "*VESTA 3 for three-dimensional visualization of crystal, volumetric and morphology data*," J. Appl. Crystallogr. **44**, 1272 (2011).

Table 1. Deviations of CMs, orientations, and molecular energies of the wall molecules at $T = 200$ K. See text for the notatins.

| Mol. No. | Position | Orientation | $\delta x$ (Å) | $\delta y$ (Å) | $\delta z$ (Å) | $\delta p$ | $\delta q$ | $\delta r$ | $\delta E$ (K) |
|---|---|---|---|---|---|---|---|---|---|
| 1 | <111> | (-111) | 0.23 | 0.11 | 0.20 | -0.08 | -0.08 | -0.02 | 69.17 |
| 2 | <11-1> | (-111) | 0.14 | 0.19 | -0.16 | 0.00 | 0.07 | -0.08 | 84.10 |
| 3 | <111> | (11-1) | 0.13 | 0.18 | 0.21 | -0.07 | -0.02 | -0.08 | 69.76 |
| 4 | <11-1> | (11-1) | 0.13 | 0.18 | -0.11 | 0.01 | -0.01 | -0.01 | 131.97 |
| 5 | <1-11> | (11-1) | 0.21 | -0.18 | 0.12 | 0.07 | -0.07 | 0.01 | 86.90 |
| 6 | <1-11> | (-111) | 0.14 | -0.14 | 0.23 | 0.04 | 0.00 | 0.04 | 32.41 |
| 7 | <1-1-1> | (-111) | 0.11 | -0.14 | -0.19 | -0.01 | 0.01 | -0.02 | 127.01 |
| 8 | <1-1-1> | (11-1) | 0.12 | -0.25 | -0.16 | 0.00 | 0.04 | 0.04 | 24.94 |
| 9 | <111> | (1-11) | 0.22 | 0.21 | 0.13 | -0.01 | -0.07 | -0.08 | 77.14 |
| 10 | <11-1> | (1-11) | 0.24 | 0.14 | -0.12 | 0.04 | 0.04 | -0.01 | 40.08 |
| 11 | <1-11> | (1-11) | 0.19 | -0.09 | 0.13 | -0.02 | -0.01 | 0.01 | 153.02 |
| 12 | <1-1-1> | (1-11) | 0.16 | -0.15 | -0.22 | -0.08 | 0.00 | 0.07 | 79.47 |
| 13 | <-111> | (1-11) | -0.18 | 0.13 | 0.19 | -0.08 | 0.00 | 0.07 | 77.03 |
| 14 | <-111> | (-111) | -0.14 | 0.12 | 0.15 | 0.00 | 0.00 | 0.00 | 150.07 |
| 15 | <-11-1> | (-111) | -0.12 | 0.16 | -0.22 | 0.04 | 0.00 | 0.04 | 34.12 |
| 16 | <-11-1> | (1-11) | -0.18 | 0.11 | -0.12 | -0.01 | 0.01 | 0.02 | 139.45 |
| 17 | <-111> | (11-1) | -0.15 | 0.23 | 0.14 | 0.01 | 0.04 | 0.05 | 34.22 |
| 18 | <-11-1> | (11-1) | -0.20 | 0.18 | -0.13 | 0.07 | -0.08 | 0.00 | 90.60 |
| 19 | <-1-11> | (1-11) | -0.23 | -0.14 | 0.13 | 0.05 | 0.05 | 0.00 | 24.28 |
| 20 | <-1-11> | (11-1) | -0.12 | -0.17 | 0.10 | 0.01 | -0.01 | 0.01 | 138.16 |
| 21 | <-1-11> | (-111) | -0.13 | -0.20 | 0.17 | 0.01 | 0.07 | -0.07 | 81.96 |
| 22 | <-1-1-1> | (-111) | -0.21 | -0.12 | -0.18 | -0.07 | -0.08 | -0.01 | 73.48 |
| 23 | <-1-1-1> | (1-11) | -0.19 | -0.22 | -0.08 | -0.02 | -0.08 | -0.07 | 55.90 |
| 24 | <-1-1-1> | (11-1) | -0.11 | -0.20 | -0.20 | -0.08 | -0.01 | -0.08 | 66.58 |

Table 2. Averaged deviations of the CM positions <*r*> and the orientations <Ω> as well as averaged STDVs of the CM positions <<*r*>> and the orientations <<Ω>> over all the wall molecules.

| Temp. | Deviation | | STDV | |
| --- | --- | --- | --- | --- |
| | <*r*> | <Ω> | <<*r*>> | <<Ω>> |
| (K) | (Å) | (deg.) | (Å) | (deg.) |
| 0 | 0.1 | 4.1 | 0.03 | 1.2 |
| 100 | 0.1 | 3.9 | 0.03 | 1.1 |
| 200 | 0.3 | 4.9 | 0.04 | 1.4 |

Table 3. Averaged deviations of the CM positions and the orientations as well as averaged STDVs of the CM positions and the orientations for the molecules in perfect crystals.

| Temp. | Deviation | | STDV | |
| --- | --- | --- | --- | --- |
| | <*r*> | <Ω> | <<*r*>> | <<Ω>> |
| (K) | (Å) | (deg.) | (Å) | (deg.) |
| 0 | 0.0 | 1.8 | 0.02 | 0.8 |
| 100 | 0.0 | 1.6 | 0.02 | 0.8 |
| 200 | 0.0 | 2.0 | 0.03 | 1.0 |

Table 4. Deviations of CMs, orientations, and molecular energies of the wall molecules forming a vacancy with a single molecule at $T = 200$ K. The notations are the same as in Table 1. See text for the notations

| Mol. No. | Position | Orientation | δx (Å) | δy (Å) | δz (Å) | δp | δq | δr | δE (K) |
|---|---|---|---|---|---|---|---|---|---|
| 1 | <111> | (-111) | 0.24 | 0.12 | 0.16 | -0.08 | -0.08 | -0.01 | 100.97 |
| 2 | <11-1> | (-111) | 0.20 | 0.11 | -0.19 | -0.01 | 0.03 | -0.05 | -253.04 |
| 3 | <111> | (11-1) | 0.14 | 0.21 | 0.21 | -0.08 | -0.01 | -0.08 | 97.49 |
| 4 | <11-1> | (11-1) | 0.14 | 0.12 | -0.14 | 0.00 | 0.01 | 0.01 | 106.30 |
| 5 | <1-11> | (11-1) | 0.25 | -0.16 | 0.13 | 0.07 | -0.07 | 0.00 | 72.81 |
| 6 | <1-11> | (-111) | 0.16 | -0.11 | 0.21 | 0.04 | 0.01 | 0.03 | 36.49 |
| 7 | <1-1-1> | (-111) | 0.23 | -0.07 | -0.11 | 0.07 | 0.05 | 0.02 | -238.85 |
| 8 | <1-1-1> | (11-1) | 0.18 | -0.20 | -0.10 | 0.00 | 0.05 | 0.05 | -53.89 |
| 9 | <111> | (1-11) | 0.22 | 0.22 | 0.10 | -0.01 | -0.07 | -0.07 | 71.85 |
| 10 | <11-1> | (1-11) | 0.22 | 0.12 | -0.19 | -0.01 | -0.01 | 0.00 | -307.62 |
| 11 | <1-11> | (1-11) | 0.22 | -0.10 | 0.10 | -0.02 | -0.02 | 0.00 | 130.40 |
| 12 | <1-1-1> | (1-11) | 0.26 | -0.14 | -0.17 | -0.10 | -0.01 | 0.07 | -402.13 |
| 13 | <-111> | (1-11) | -0.18 | 0.14 | 0.22 | -0.07 | 0.01 | 0.07 | 67.22 |
| 14 | <-111> | (-111) | -0.11 | 0.13 | 0.18 | 0.01 | 0.02 | -0.01 | 156.26 |
| 15 | <-11-1> | (-111) | -0.11 | 0.12 | -0.23 | 0.04 | 0.00 | 0.03 | 39.14 |
| 16 | <-11-1> | (1-11) | -0.17 | 0.12 | -0.12 | -0.02 | 0.00 | 0.02 | 156.57 |
| 17 | <-111> | (11-1) | -0.14 | 0.22 | 0.13 | 0.00 | 0.03 | 0.04 | 50.25 |
| 18 | <-11-1> | (11-1) | -0.19 | 0.15 | -0.13 | 0.07 | -0.08 | 0.00 | 97.25 |
| 19 | <-1-11> | (1-11) | -0.24 | -0.11 | 0.16 | 0.05 | 0.05 | 0.00 | 34.92 |
| 20 | <-1-11> | (11-1) | -0.10 | -0.15 | 0.11 | 0.01 | -0.02 | -0.01 | 172.87 |
| 21 | <-1-11> | (-111) | -0.12 | -0.17 | 0.16 | 0.01 | 0.07 | -0.07 | 72.23 |
| 22 | <-1-1-1> | (-111) | -0.20 | -0.08 | -0.19 | -0.08 | -0.07 | -0.02 | 107.23 |
| 23 | <-1-1-1> | (1-11) | -0.18 | -0.20 | -0.11 | -0.02 | -0.08 | -0.07 | 98.19 |
| 24 | <-1-1-1> | (11-1) | -0.09 | -0.17 | -0.18 | -0.08 | -0.02 | -0.08 | 100.07 |

Table 5. Averaged deviations of the CM positions $<r>$ the orientations $<\Omega>$ as well as averaged STDVs of the CM positions $<<r>>$ and the orientations $<<\Omega>>$ over the wall molecules forming a vacancy with a single $CO_2$ molecule.

| Temp. | Deviation | | STDV | |
|---|---|---|---|---|
| | $<r>$ | $<\Omega>$ | $<<r>>$ | $<<\Omega>>$ |
| (K) | (Å) | (deg.) | (Å) | (deg.) |
| 0 | 0.1 | 3.9 | 0.02 | 0.6 |
| 100 | 0.1 | 4.1 | 0.03 | 1.2 |
| 200 | 0.3 | 4.9 | 0.04 | 1.4 |

Table 6. Averaged STDVs of the CM positions $<<r>>$ and the orientations $<<\Omega>>$ of the single $CO_2$ molecule in the vacancy.

| Temp. | STDV | |
|---|---|---|
| | $<<r>>$ | $<<\Omega>>$ |
| (K) | (Å) | (deg.) |
| 0 | 0.02 | 0.7 |
| 100 | 0.04 | 1.2 |
| 200 | 0.05 | 1.7 |

*Figure captions*

Fig. 1. Configurations of the wall molecules at *T* = 0 K observed from the [111] direction of the *Pa3* crystal.

Fig. 2. Configurations of the wall molecules and the $CO_2$ molecule (light-blue) added in the cubic vacancy at *T* = 200 K observed from the [111] direction of the *Pa3* crystal.

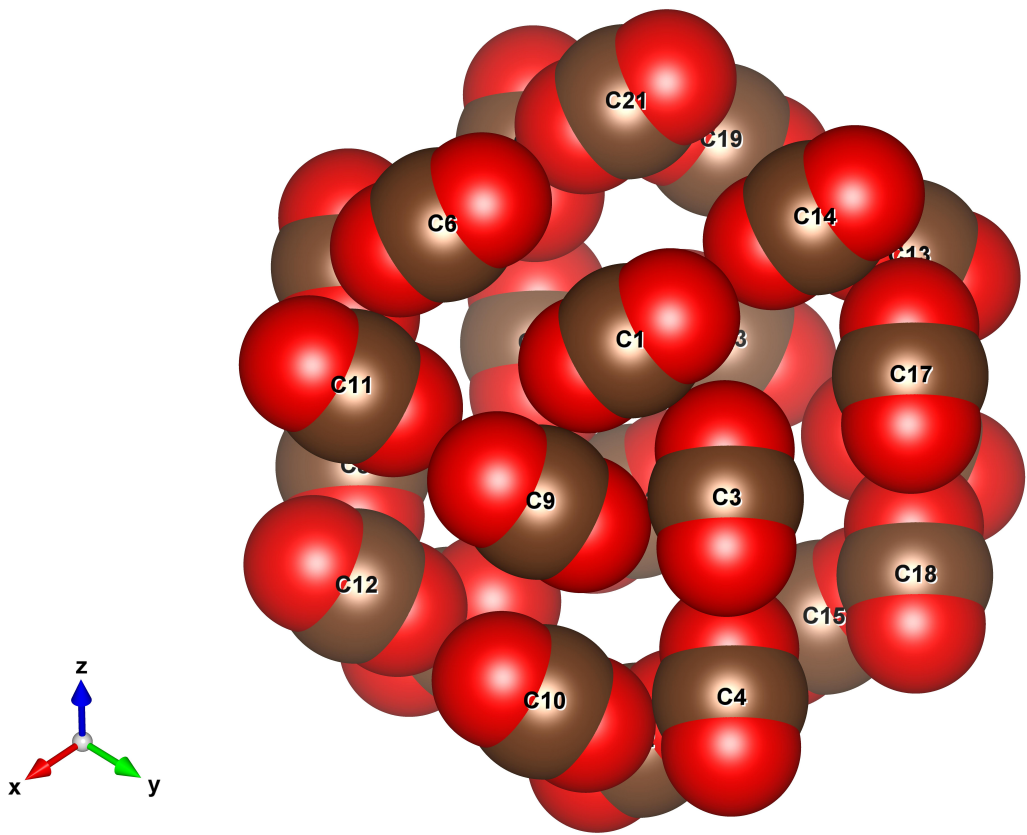

Fig. 1. Configurations of the wall molecules at $T = 0$ K observed from the [111] direction of the *Pa3* crystal.

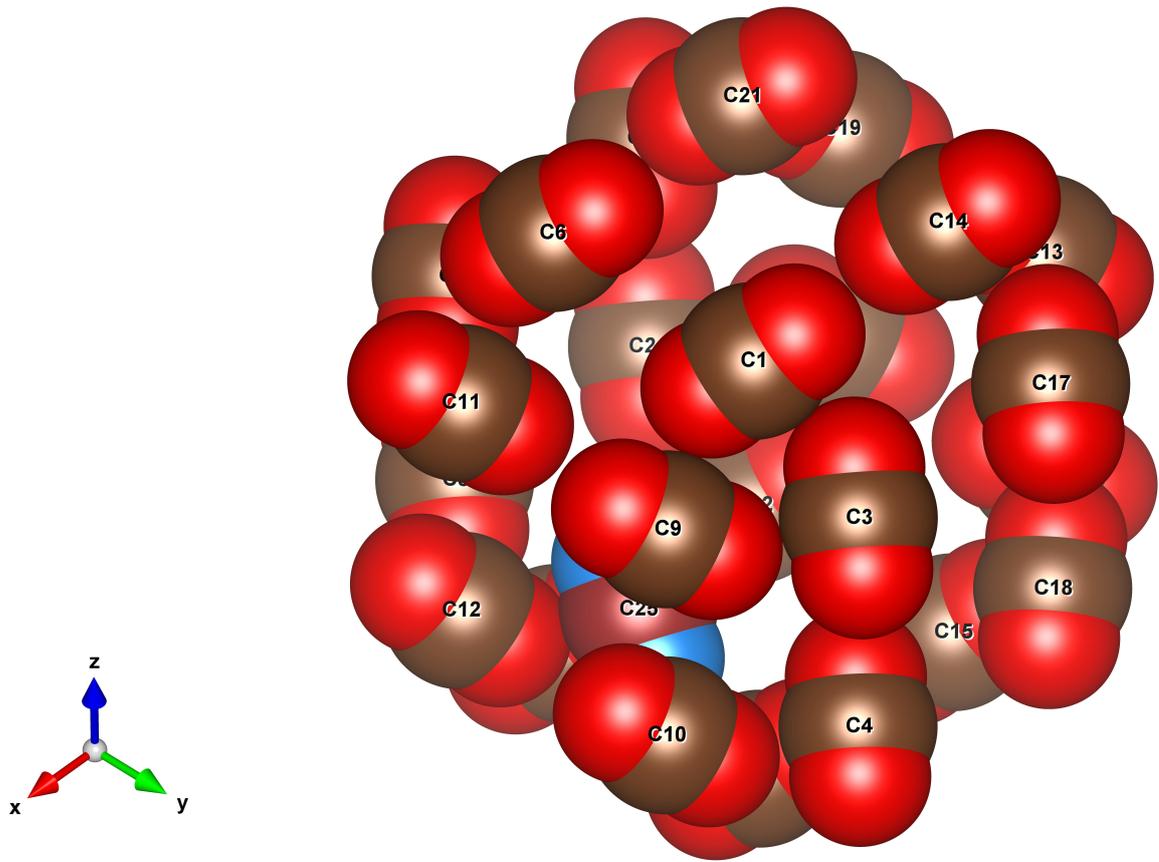

Fig. 2. Configurations of the wall molecules and the $CO_2$ molecule (light-blue; molecule 25) added in the cubic vacancy at $T = 200$ K observed from the [111] direction of the $Pa3$ crystal.